\documentstyle[aps,epsf,multicol,psfig]{revtex}
\begin{document}
\date{June 26, 2001}
\title{Phonon spectrum in a nanoparticle mechanically coupled to a substrate}
\author{Kelly R. Patton and Michael R. Geller}
\address{Department of Physics and Astronomy, University of Georgia, Athens, 
Georgia 30602-2451, USA}
\maketitle

\begin{abstract}
We calculate the vibrational density-of-states in an insulating
nanoparticle that is in weak mechanical contact with a semi-infinite 
substrate. The work is motivated by a recent experiment by Yang {\it et al.}, 
where the low-energy phonon density-of-states of ${\rm Y_2 O_3}$ nanoparticles
doped with Eu$^{3+}$ was measured. Preliminary results presented here, based 
on the conventional quasiparticle-pole approximation for the phonon 
propagator, are in reasonable agreement with experiment.
\end{abstract}


\begin{multicols}{2}

\section{introduction}

As is well known, the vibrational spectrum of an isolated nanometer-scale 
crystal is discrete. For a spherical nanoparticle of diameter $d$ and bulk 
transverse sound velocity $v_{\rm t}$, the lowest internal vibrational mode, 
called the Lamb mode, has a frequency of about $2\pi v_{\rm t}/d$. Below this 
frequency no internal vibrational modes exist, and the phonon 
density-of-states (DOS) vanishes. Any property of the nanoparticle that 
depends on this vibrational spectrum, such as its thermodynamic properties or 
electron-phonon dynamics, will therefore be very different, especially at low 
energies, when compared to bulk crystals made of the same material. 

In an interesting recent experiment, Yang {\it et al.} \cite{Yang etal} 
measured the DOS of ${\rm Y_2 O_3}$ nanoparticles with a size distribution 
ranging from 7 to 23 nm in diameter. Nanoparticles of size, say, 15 nm, cannot
support phonons with energies below about 10 ${\rm cm^{-1}}$. The phonon DOS 
was obtained by measuring the nonradiative lifetime of an excited electronic 
state of Eu$^{3+}$, and at 3 ${\rm cm^{-1}}$ was found to be about 100 times 
smaller than that of a bulk ${\rm Y_2 O_3}$ crystal.

In this paper we propose and investigate a mechanism to explain the observed
low-energy DOS. Several phonon-broadening mechanisms could be responsible for 
this effect. For example, anharmonicity causes the modes to broaden, leading to
a DOS at low energy. However, anharmonicity is ineffective at low energy and 
an estimate of the anharmonic broadening showed that the resulting DOS is 
much smaller than that observed\cite{Markel and Geller}. Adsorbed molecules or
``dirt'' on the outside of the nanoparticle could also be involved. 
A more likely broadening mechanism results from the fact that these 
nanoparticles are not isolated, but rather are in contact with each other or 
some support structure. This contact enables the nanoparticles to couple to an
environment with a continuous spectrum at low energy. We believe that it is 
this physical coupling to the environment that broadens the vibrational modes 
enough to explain the observed DOS. 

Our preliminary results, presented here, are favorable. However, the 
calculation has only been done within the conventional quasiparticle-pole 
approximation for the nanoparticle phonon propagator\cite{Abrikosov etal}, 
which, to leading order in perturbation theory, is equivalent to using Fermi's
golden rule to obtain vibrational-mode lifetimes and assuming Lorentzian 
line shapes. A more accurate calculation of the nanoparticle DOS, which 
requires the solution of the Dyson equation, will be presented in future work.

\section{model and method}   

Our model consists of a single isotropic elastic sphere of diameter 10 nm, 
representing the nanoparticle, connected to a semi-infinite isotropic elastic 
continuum lying in the $x y$ plane and extending to infinity in the negative 
$z$ direction. For simplicity, we take the substrate and the nanoparticle to 
be made of the same material. We model the contact between the two by a weak 
harmonic spring, corresponding to the situation where the nanoparticle and
substrate are connected by only a few atomic bonds or by a small ``neck'' of 
material. The Hamiltonian for the system is
\begin{eqnarray}
H &=& \sum_{J}\hbar \omega_{J}a_{J}^{\dag}a_{J}+\sum_{I}\hbar \omega_{I}
b_{I}^{\dag}b_{I} \nonumber \\
&+& \frac{1}{2} K \big[u_{z}({\bf r}_0)_{\rm part} - u_{z}({\bf r}_0)_{\rm
sub} \big]^2,
\label{hamiltonian}
\end{eqnarray}
where the $a_J$ and $a_J^{\dag}$ are annihilation and creation operators for 
phonons in the nanoparticle, and the index $J$ runs over all the modes of the 
nanoparticle. The $b_I$, $b_I^{\dag}$, and $I$ correspond to the substrate 
phonons \cite{substrate footnote}. The spring constant $K$ is taken to be of 
the order of an atomic bond strength of the material. $u_{z}({\bf r}_0)_{\rm
part}$ and $u_{z}({\bf r}_0)_{\rm sub}$ are the $z$ components of the phonon 
displacement field of the nanoparticle and the substrate respectively, 
evaluated at the point of connection ${\bf r}_0$. 

The displacement field for the nanoparticle can be expanded as
\begin{equation}
{\bf u}({\bf r},t)_{\rm part} = \sum _{J}\sqrt{\frac{\hbar}{2\rho 
\omega_{J}}} \big[ a_{J}{\bf \Psi}_{J}({\bf r})e^{-i\omega_{J}t} + {\rm H.c.}
\big],
\end{equation}
where the ${\bf \Psi}_{J}({\bf r})$ are the nanoparticle vibrational 
eigenmodes, normalized according to
\begin{equation}
\int_{V} d^3r \ {\bf \Psi}^{*}_J({\bf r})\cdot{\bf \Psi}_{J'}({\bf r}) = 
\delta_{JJ'}, 
\end{equation}
and $\rho$ is the mass density. $V$ is the volume of the nanoparticle. The 
nanoparticle's vibrational eigenmodes were found by using a method similar to 
the one used by Lamb\cite{Lamb}.

Similarly, the substrate displacement field is given by
\begin{equation}
{\bf u}({\bf r},t)_{\rm sub} = \sum _{I}\sqrt{\frac{\hbar}{2\rho \omega_{I}}}
\big[ b_{I}{\bf f}_{I}({\bf r})e^{-i\omega_{I}t} + {\rm H.c.} \big],
\end{equation}
with the ${\bf f}_I$ normalized as above. The vibrational eigenmodes of the
substrate were calculated following Ezawa{\cite{Ezawa}}.

To determine the vibrational DOS we calculate the retarded Green's function 
for the nanoparticle, 
\begin{equation}
D^{ij}({\bf r},{\bf r}',t) \equiv -i\theta(t)\left<\left[u^{i}({\bf r},t),u^{j}
({\bf r}',0)\right]\right>.
\end{equation}
Evaluating this perturbatively to second order in the spring constant $K$, and 
using the quasiparticle-pole approximation\cite{Abrikosov etal} (which 
replaces the energy-dependent phonon self-energy for each mode $J$ with it's 
on-shell value), we find the energy damping rate for mode $J$ to be given by 
\begin{equation}
\tau^{-1}_{J} = \frac{\pi K^2}{\hbar \rho}\frac{N_{\rm s}(\omega_{J})}
{\omega_{J}} \, \big|\Psi^{z}_{J}({\bf r}_{0}) \big|^2,
\label{damping rate}
\end{equation}
where 
\begin{equation}
N_{\rm s}(\omega) \equiv - {1 \over \pi} \, {\rm Im} \, D_{\rm s}^{zz}
({\bf r}_0, {\bf r}_0, \omega)
\label{spectral density definition}
\end{equation}
is the phonon spectral density at the surface of the substrate and $D_{\rm 
s}^{ij}({\bf r},{\bf r}',\omega)$ is the transform of the substrate 
propagator. Eq.~(\ref{damping rate}) can also be derived from Fermi's golden 
rule. The quasiparticle-pole approximation predicts Lorentzian-broadened line 
shapes. This leads to a nanoparticle DOS given by
\begin{equation}
g(\epsilon) = \sum_{J}\frac{\hbar \tau^{-1}_{J}/\pi}{(\epsilon - \hbar 
\omega_{J})^2 + \hbar^2 \tau_{J}^{-2}}.
\label{nanoparticle DOS definition}
\end{equation}
The usual thermodynamic DOS (number of states per unit energy per unit volume)
is given by $g(\epsilon)/V$, where $V$ is again the volume of
the nanoparticle.

\section{results and conclusions}

\begin{figure}
\centerline{\psfig{file=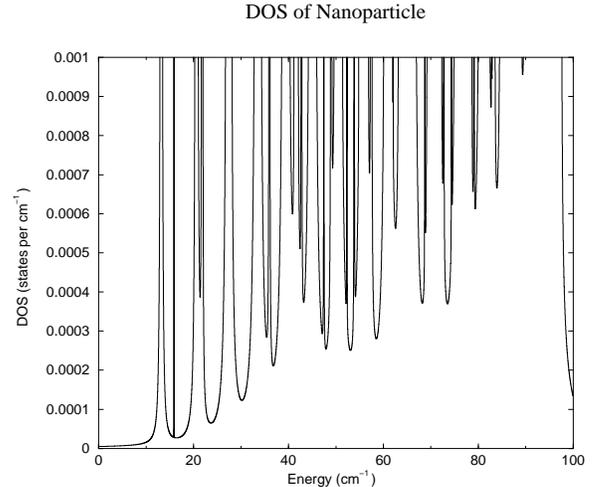,width=3.0in,angle=-90}}
\vspace{0.1in}\setlength{\columnwidth}{3.2in}
\centerline{\caption{Vibrational DOS in a 10 nm Si nanoparticle, weakly 
coupled to a substrate. The DOS rapidly approaches the bulk Debye law.
\label{high energy dos figure}}}
\end{figure}

For simplicity we assume the nanoparticle and substrate to be made of Si, thus 
enabling us to obtain the surface spectral density (\ref{spectral density 
definition}) from Appendix B of Ref.~\onlinecite{Patton and Geller thermal 
tunneling}. We treat Si as an isotropic elastic continuum with longitudinal 
and transverse sound velocities
\begin{eqnarray}
v_{\rm l} &=& 8.5 \times 10^5 \, {\rm cm \ s}^{-1}, \nonumber \\
v_{\rm t} &=& 5.9 \times 10^5 \, {\rm cm \ s}^{-1},
\label{Si velocities}
\end{eqnarray}
and mass density $\rho = 2.3  \, {\rm g \ cm}^{-3}.$ For a spring stiffness 
$K$ equal to $1.0 \times 10^{4} \, {\rm erg \ cm^{-2}}$, we obtain the phonon 
DOS shown in Fig.~\ref{high energy dos figure}.

Our calculation includes all modes below a cutoff frequency $\omega_{\rm max}$
of 100 ${\rm cm}^{-1}$. The low-energy DOS depends, to some extent, on our
choice of $\omega_{\rm max}$. This sensitivity is an artifact of the 
quasiparticle-pole approximation, or equivalently, a consequence of assuming
Lorentzian line shapes. Because at this stage we are only interested in
whether our result agrees with the experiment of Ref.~\cite{Yang etal} at the 
order-of-magnitude level, we will not discuss the weak $\omega_{\rm max}$ 
dependence any further.

\begin{figure}
\centerline{\psfig{file=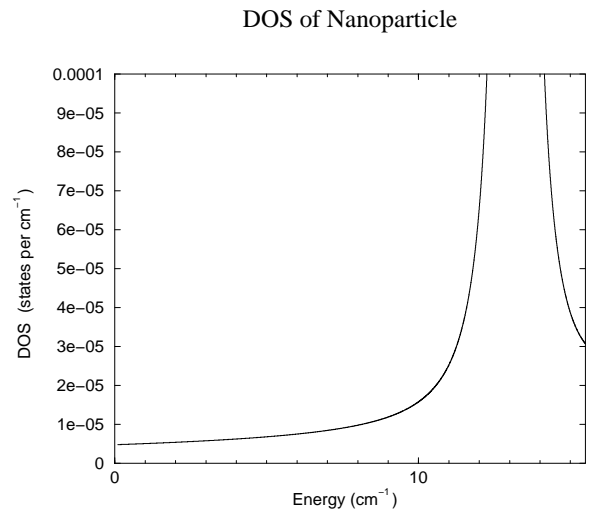,width=3.0in}}
\vspace{0.1in}\setlength{\columnwidth}{3.2in}
\centerline{\caption{Low-energy DOS.\label{low energy dos figure}}}
\end{figure}

Fig.~\ref{low energy dos figure} shows the low-energy DOS up to about 15 
${\rm cm}^{-1}$. The large peak on the right is the five-fold-degenerate
Lamb mode. At 3 ${\rm cm}^{-1}$ the DOS is found to be $5.8 \times 10^{-6}
\ {\rm states \ per \ cm^{-1}}$, or about $3.2 \times 10^{-4}$ times the bulk 
DOS \cite{bulk DOS footnote}
\begin{equation}
\sum_\lambda {\epsilon^2 \over 2 \pi^2 \hbar^3 v_\lambda^3} \cdot V = 2.0 
\times 10^{-3} \, E^2 \ { {\rm states} \over {\rm cm}^{-1}}, 
\label{bulk DOS}
\end{equation}
where $E$ is the energy in cm$^{-1}$. Experimentally, the ratio of nanoparticle
to bulk DOS was found to be approximately $7.4 \times 10^{-3}$, about 20 times
larger than our result.

This is a good result considering the simplicity of our model. However, the 
quasiparticle-pole approximation is probably inaccurate much below
10 cm$^{-1}$, where the deviation from Lorentzian line shapes becomes 
important. However, the order-of-magnitude agreement does suggest that we have 
correctly identified the relevant broadening mechanism in these nanoparticles.
An accurate calculation of the nanoparticle DOS, based on the solution of the 
Dyson equation, is in progress.

\section{acknowledgements}

This work was supported by the National Science Foundation under CAREER 
Grant No.~DMR-0093217, and by a Research Innovation Award and Cottrell 
Scholars Award from the Research Corporation. It is a pleasure to thank
Bill Dennis and Richard Meltzer for useful discussions, and Patrick 
Sprinkle for help with the numerics.

\end{multicols}


\begin{references}

\bibitem{Yang etal} H. S. Yang, S. P. Feofilov, D. K. Williams, J. C. Milora, 
B. M. Tissue, R. S. Meltzer, and W. M. Dennis, Physica B {\bf 263}, 476 (1999).

\bibitem{Markel and Geller} V. A. Markel and M. R. Geller, J. Phys. Condens. 
Matter {\bf 12}, 7569 (2000).

\bibitem{Abrikosov etal} A. A. Abrikosov, L. P. Gorkov, and I. E. 
Dzyaloshinski, {\it Methods of Quantum Field Theory in Statistical Physics}
(Dover, New York, 1975), p. 60.

\bibitem{substrate footnote} Because the substrate is semi-infinite, the label
$I$ assumes both discrete and continuous values.

\bibitem{Lamb} H. Lamb, Proc. London Math. Soc. {\bf 13}, 187 (1882).

\bibitem{Ezawa} H. Ezawa, Ann. Phys. {\bf 67}, 438 (1971).

\bibitem{Patton and Geller thermal tunneling} K. R. Patton and M. R. Geller, 
cond-mat/0101045, to appear in Phys. Rev. B.

\bibitem{bulk DOS footnote} The standard Debye formula has to be multiplied by
$V$ to to make comparison to Eq.~(\ref{nanoparticle DOS definition}) possible.

\end{references}
\end{document}